# Thermopower analysis of metal-insulator transition temperature modulations in vanadium dioxide thin films with lattice distortion

Takayoshi Katase[*], Kenji Endo, and Hiromichi Ohta[*]

*Research Institute for Electronic Science, Hokkaido University, N20W10, Sapporo 001-0020, Japan*

[*] Correspondence to: katase@es.hokudai.ac.jp, hiromichi.ohta@es.hokudai.ac.jp

Insulator-to-metal (MI) phase transition in vanadium dioxide ($VO_2$) thin films with controlled lattice distortion was investigated by thermopower measurements. $VO_2$ epitaxial films with different crystallographic orientations, grown on (0001) α-$Al_2O_3$, (11$\bar{2}$0) α-$Al_2O_3$, and (001) $TiO_2$ substrates, showed significant decrease of absolute value of Seebeck coefficient ($S$) from ~200 to 23 μV $K^{-1}$, along with a sharp drop in electrical resistivity ($\rho$), due to the transition from an insulator to a metal. The MI transition temperatures observed both in $\rho$ ($T_\rho$) and $S$ ($T_S$) for the $VO_2$ films systematically decrease with lattice shrinkage in the pseudo-rutile structure along $c$-axis, accompanying a broadening of the MI transition temperature width. Moreover, the onset $T_S$, where the insulating phase starts to become metallic, is much lower than onset $T_\rho$. This difference is attributed to the sensitivity of $S$ for the detection of hidden metallic domains in the majority insulating phase, which cannot be detected in $\rho$-measurements. Consequently, $S$-measurements provide a straightforward and excellent approach for a deeper understanding of the MI transition process in $VO_2$.

# I. INTRODUCTION

Vanadium dioxide ($VO_2$), a strongly correlated oxide, shows an insulator-to-metal (MI) transition when heated to its transition temperature ($T_{MI}$) of 68 °C[1] due to crystal-structure and electronic-structure changes. In the high-temperature metallic phase at $T > T_{MI}$, $VO_2$ has a rutile-type tetragonal structure, while in the low-temperature insulating one at $T < T_{MI}$, the vanadium ions form a dimer along with the position shifting from linear chains to zigzag pattern, resulting in a monoclinic structure.[2] This structural transformation accompanies a dramatic change in the 3d-band configuration with appearance of bandgap ~0.6 eV,[3] yielding abrupt changes of both electrical resistivity in the order of $10^4$–$10^5$ and the optical transmission in the infrared region. $T_{MI}$ can be varied over a wide range below room temperature by aliovalent ion doping,[4] external strain,[5] and an electric field application.[6] These features of the MI transition in $VO_2$ have proved to be useful in electrical and optical switching devices.[7,8]

However, the nature of the MI transition in $VO_2$ is still quite unclear[9,10] and consequently potential device applications remain underdeveloped. One reason for this situation is that $VO_2$ has an electronic phase separation at the nanoscale region[11] and a complicated phase diagram with internal strain[12], which significantly makes it difficult to understand and control the MI transition characteristics. In a temperature range around the $T_{MI}$, the metallic tetragonal and insulating monoclinic phases coexist randomly, where the electrical-transport properties are largely influenced by the domain configuration.[13] Especially in thin films, the polycrystalline domains with epitaxial strains and defects cause discontinuous and irregular phase domain structures, which result in the complex and broad MI transition.[14]

The phase domain structures and their electronic behaviors have been experimentally investigated by electrical-transport measurement using nanowires at the nanoscale[15] or nanoscopic imaging on thin-films during the MI transitions.[16] In the electrical-transport measurement on $VO_2$ nanowires, discrete resistivity jump was observed, which suggests the important role of the electronic behaviors of each domain in the electrical properties. The nanoscopic imaging on thin films revealed that both electronic and structural changes exhibit phase coexistence leading to percolation conduction depending on the domain distribution. It is therefore critically important to understand the evolution of coexisting metallic and insulating domains with spatial phase inhomogeneity at the nano-scale and further characterization that combines several experimental techniques is necessary to unveil and control the MI transition process.

We have focused on thermopower measurement as a physical property to investigate electronic structure change around the MI transition of $VO_2$,[17] because Seebeck coefficient ($S$) changes significantly due to the electronic-structure reconstruction across the $T_{MI}$. The general expression for the electronic contribution to the $S$ is

$$S = -\frac{k_B}{e} \int (\frac{E-E_F}{k_B T}) \frac{\sigma(E)}{\sigma} dE \quad (1)$$

, where $k_B$ is Boltzmann constant and $E_F$ is Fermi energy.[18] Here, $\sigma$ is described as an integral over the single carrier without electron correlation effects,

$$\sigma = \int \sigma(E) dE = e \int g(E) \mu(E) f(E)[1-f(E)] dE \quad (2)$$

, where $g(E)$ is the density of state (DOS), $\mu(E)$ is the mobility, $f(E)$ is the Fermi distribution function, and $\sigma(E)$ is energy-dependent conductivity.

$S$ of semiconductors is expressed as

$$S = \frac{k_B}{e}(\frac{E_F - E_c}{k_B T} + A) \quad (3)$$

, considering (1) and (2) in the framework that only electrons contribute to the $S$ and $E_F$ lies near the conduction band edge ($E_c$). By using the relation of $n_e = N_c e^{\frac{E_F - E_C}{k_B T}}$ in (3), the $S$ can be simplified to

$$S = -\frac{k_B}{e}(\ln \frac{N_c}{n_e} + A) \quad (4)$$

, where $N_c$ is effective DOS of the conduction band, $n_e$ is carrier concentration, and $A$ is a transport constant that depends on the dominant scattering mechanism.[19] Meanwhile, $S$ for metals or degenerate semiconductors at low temperature ($E_F \gg$ thermal energy $k_B T$) is basically expressed as

$$S = \frac{\pi^2}{3} \frac{k_B^2 T}{e} \left\{ \frac{\partial [\log(\sigma(E))]}{\partial E} \right\}_{E=EF} \quad (5)$$

in Mott's equation.[20,21]

$S$ of metals are typically small and keep constant by reflecting the energy differential of DOS around the $E_F$, but those of the semiconductors drastically change, depending on $n_e$, by reflecting the shape of DOS around conduction band bottom due to the $E_F$ shifts by carrier doping;[22] the doped carriers may be unintentionally caused by the existence of oxygen vacancies for $VO_2$ thin film.[23] Previously, we systematically measured the $S$ of electron-doped $V_{1-x}W_xO_2$ epitaxial films with different doping levels,[17] and clarified that the metallic electronic structure clearly changes into an insulating electronic structure with a parabolic DOS around the conduction band bottom, where the $S$ for insulating phase systematically changes by reflecting the derivative of the DOS around the $E_F$.

$S$ measurements have been actively performed to investigate the MI transition characteristics of bulks,[24,25] nanowires,[26] and thin-films[27] of VO$_2$, where the $S$ significantly change from 200–400 µV K$^{-1}$ for the insulating phase down to ~−20 µV K$^{-1}$ for the metallic phase. In addition, $S$ measurements were recently applied to investigate the metal-insulator domain arrays in single-crystal microbeams[26] and the effect of the domain inhomogeneity in thin-films,[27] which demonstrates the unusual $S$; the complications of the phase coexistence near the domain wall may cause a deviation of the $S$ from a normal behavior in each pure domain. The ability of sensitive detection of the metallic and insulating phases around the $T_{MI}$ should make $S$-measurements suitable to investigate the MI transition process of VO$_2$.

In this paper, we investigate the insulator-to-metal phase transition of VO$_2$ thin films with regulated lattice distortion via $S$-measurements. We controlled the epitaxial orientations of the VO$_2$ films grown on single crystalline substrates of (0001) α-Al$_2$O$_3$, (11$\bar{2}$0) α-Al$_2$O$_3$, and (001) rutile-TiO$_2$. Then we compared their MI transition process by simultaneously measuring the $S$ and electrical resistivity ($\rho$).

**II. EXPERIMENTAL**

VO$_2$ epitaxial films were grown on (0001) α-Al$_2$O$_3$, (11$\bar{2}$0) α-Al$_2$O$_3$, and (001) TiO$_2$ substrates by pulsed laser deposition. A KrF excimer laser with a laser energy fluence of 3 J cm$^{-2}$ was used to ablate a V$_2$O$_5$ target disk. The growth temperature was fixed at 500 °C, and the oxygen pressure ($P_{O2}$) was optimized at 2.0 Pa because the resistivity-change ratio from an insulator to a metal across the $T_{MI}$ is extremely sensitive to $P_{O2}$ during thin film growth.[28] After deposition at a rate of approximately 2 nm min$^{-1}$, the films were cooled to room temperature (RT) under the same oxygen pressure. The film thickness was set at 30–40 nm, which was characterized by X-ray reflectivity measurements. The crystal structures, including the crystallographic orientation and the mosaicity of the crystallites, were investigated by high-resolution X-ray diffraction (XRD, anode radiation: monochromatic CuK$\alpha_1$, ATX-G, Rigaku Co.) at RT. The film surface structures were observed by atomic force microscopy (AFM).

$S$ was measured by giving a temperature difference ($\Delta T$) of ~4 K in the film with In-Ga alloy electrodes. Here we used two Peltier devices placed under the film, where one side cooled and the other side heated a sample, without changing the temperature at the center.[29] We used two Cu leads to measure the thermo-electromitive force ($\Delta V$). Thus, observable slope of $\Delta T/\Delta V$ plots shoud be expressed as $S_{VO2\ film} + S_{Cu}$ [see Supplementary Material for the calculation of $S_{VO2\ film} + S_{Cu}$]. Since $S$ of Cu is very small (~+1.5 μV K$^{-1}$) as compared to that of VO$_2$, $S$ of the Cu leads was ignored in this study. The actual temperatures of both sides of the VO$_2$ film surface were monitored by two tiny thermocouples separated by a gap ~7 mm. The thermo-electromotive force ($\Delta V$) and $\Delta T$ were simultaneously measured at almost the same positions, and the $S$ were obtained from the linear slope of the $\Delta V$–$\Delta T$ plots.[17] It should be noted that the sapphire substrate is not suited for $S$-measurements because of high-thermal conductivity, but we measured the $\Delta V$ at the highly-steady state after making the $\Delta T$ up to 10 K in sapphire substrate, which enables accurate $S$-measurements of VO$_2$ thin-films prepared on sapphire substrates. Meanwhile, $\rho$ was measured by d.c. four probe method using Au electrodes.

## III. RESULTS

### A. Epitaxial orientation control of VO$_2$ thin films

Figure 1 shows the out-of-plane (a–c) and in-plane (d–f) XRD patterns of VO$_2$ films grown on the substrates of (0001) α-Al$_2$O$_3$ (a, d), (11$\bar{2}$0) α-Al$_2$O$_3$ (b, e), and (001) TiO$_2$ (c, f). In addition, Fig. 2 summarizes (a–c) AFM images and reflection high-energy electron diffraction (RHEED) patterns, and (d–f) schematic epitaxial relation models of each film. For the out-of-plane XRD pattern of VO$_2$ film grown on (0001) α-Al$_2$O$_3$ substrate (Fig. 1(a)), diffraction peaks of 0$k$0 ($k$ = 2 and 4) for monoclinic VO$_2$ are observed along with those of the (0001) α-Al$_2$O$_3$ substrate. The average crystalline tilt angle, full width at the half maximum (FWHM) value of the 020 diffraction peak (2$\theta$-fixed $\omega$-scan, inset of Fig. 1(a)), is 0.17°. For the in-plane XRD pattern (Fig. 1(d)), 200 VO$_2$ and 11$\bar{2}$0 α-Al$_2$O$_3$ diffraction peaks are observed, which confirms the epitaxial relationship of (010)[100] VO$_2$ ∥ (0001)[11$\bar{2}$0] α-Al$_2$O$_3$. These results are also consistent with the RHEED pattern (Fig. 2(a)). However, the in-plane 2$\theta_\chi$-fixed $\phi$-scan of the 200 diffraction peak (inset of Fig. 1(d)) shows six diffraction peaks instead of the two-fold rotational symmetry of a monoclinic lattice, indicating that the VO$_2$ film has three domain structures with 60° rotational periodicity, presumably due to domain matching between the film and substrate. Meanwhile, the broad 200 diffraction peaks, which indicate a weak in-plane orientation, should originate from the misaligned orientations between (100) VO$_2$ and (11$\bar{2}$0) α-Al$_2$O$_3$ by 2.1° due to the monoclinic structure, as shown in the schematic epitaxial relation model (Fig. 2(d)). Reflecting the rotational domain structure, small crystalline grains with size of ~20 nm are observed in the topographic AFM image (Fig. 2(a)).

On the other hand, in the case of VO$_2$ film on (11$\bar{2}$0) α-Al$_2$O$_3$ substrate, $h$00 ($h$ = 2, 3, and 4) diffraction peaks of monoclinic VO$_2$ are observed along with intense peaks from the (11$\bar{2}$0) α-Al$_2$O$_3$ substrate in the out-of-plane XRD pattern (Fig. 1(b)). Because the double lattice spacing ($h$ = odd numbers) along the $a$-axis of monoclinic VO$_2$ originates from the formation of the vanadium-ion dimer, the existence of 300 diffraction peak confirms that the structural transition temperature (monoclinic to tetragonal phase transition) is above RT. The average crystalline tilt angle is 0.16° (inset of Fig. 1(b)). The in-plane XRD pattern (Fig. 1(e)) shows clear diffraction peaks of 020 VO$_2$ and 0006 α-Al$_2$O$_3$. The in-plane $\phi$ scan of the 020 diffraction shows a two-fold rotational symmetry with 180°, originating from the monoclinic symmetry of the VO$_2$ lattice. These results substantiate that the VO$_2$ film is grown on the (11$\bar{2}$0) α-Al$_2$O$_3$ substrate with an epitaxial relationship of (100)[010] VO$_2$ ∥ (11$\bar{2}$0)[0001] α-Al$_2$O$_3$ (Fig. 2(e)), which is also confirmed in the RHEED pattern (Fig. 2(b)). For film surface structure of the VO$_2$ film on (11$\bar{2}$0) α-Al$_2$O$_3$ substrate, reflecting the anisotropic crystal structure of VO$_2$ along the [102] direction, rectangular shaped grains are observed in the AFM image (Fig. 2(b)).

For out-of-plane XRD pattern of VO$_2$ film grown on (001) TiO$_2$ substrate (Fig. 1(c)), the $40\bar{2}$ diffraction peak of VO$_2$ is observed along with the 002 diffraction peak of TiO$_2$. The $40\bar{2}$ diffraction corresponds to 002 diffraction of tetragonal VO$_2$, which has the same rutile structure with the TiO$_2$ substrate. For the in-plane XRD pattern (Fig. 1(f)), the 022 diffraction of monoclinic VO$_2$, corresponding to the 220 diffraction peak of tetragonal VO$_2$, overlaps with the 220 diffraction peak of TiO$_2$ substrate. Asymmetric $\phi$-scan for the 420 (monoclinic) [222 (tetragonal)] diffraction peak, shown in the inset, exhibits a four-fold rotational symmetry with 90°, confirming that VO$_2$ film grows on isostructural TiO$_2$ (001) substrates with an epitaxial relationship between the film and TiO$_2$ substrate [*i.e.* ($20\bar{1}$)[011] VO$_2$ (monoclinic) ∥ (001)[110] TiO$_2$ or (001)[110] VO$_2$ (tetragonal) ∥ (001)[110] TiO$_2$ (Fig. 2(f))]. For film surface of the VO$_2$ film on the (001) TiO$_2$ substrate (Fig. 2(c)), the film has a polycrystalline structure composed of small grains (~40 nm diameters), although epitaxial growth is confirmed in the RHEED patterns. From these results, we successfully controlled the crystal orientations of the VO$_2$ films by the choice of substrates.

**B. Metal-insulator transition characteristics**

Figures 3(a) and (b) summarize $\rho-T$ curves normalized by $\rho$ at 35 °C ($\rho/\rho_{35°C}$) and $S-T$ curves of the VO$_2$ epitaxial films, respectively, where it should be noted that there has been no report on the comparison of the MI transition characteristics for VO$_2$ films with different crystallographic orientations. Here, we normalized $\rho$ to show the change of $T_{MI}$ clearly, because $\rho$ of VO$_2$ epitaxial films were scattered as shown in the inset of Fig.3(a), where the $\rho$ should be influenced by grain-boundary scattering due to different grain sizes and different epitaxial orientations of the VO$_2$ films. The average $\rho_{35°C}$ was 1.95 Ω cm with a standard deviation of 1.12 Ω cm. The average $\rho$ was confirmed to be consistent with previously reported values of VO$_2$ films.[27] Meanwhile, $S$ is a physical property that is not affected by grain boundary scattering and make it easy to directly compare MI transition characteristics. Both $\rho$–$T$ and $S$–$T$ were measured along the [100], [010], and [011] directions for monoclinic VO$_2$ films grown on (0001) α-Al$_2$O$_3$, (11$\bar{2}$0) α-Al$_2$O$_3$, and (001) TiO$_2$, respectively. As the temperature increases, all the VO$_2$ films show a sharp drop (a three-digit decrease) in $\rho$–$T$ (Fig. 3(a)) due to the transition from an insulator to a metal. For the $S$–$T$ curves (Fig. 3(b)), |$S$| of all the VO$_2$ films start to decrease from 220–240 μV K$^{-1}$ for the insulating phase down to 23 μV K$^{-1}$ for the metallic phase.[17] The |$S$| of insulating phase at low temperature around RT increases linearly with temperature, indicating that the insulating phase is a degenerate semiconductor, while the constant $S$ for the metallic phase (−23 μV K$^{-1}$) agree well with the previously reported $S$ of ~−20 μV K$^{-1}$ for those of undoped VO$_2$ bulks,[24,25] microbeams,[26] and thin films.[27]

To visualize the MI transitions clearly, Figs. 3(c) and (d) plot the temperature derivative curves of d[log $\rho$ / $\rho_{35°C}$] / d$T$ and d$S$ / d$T$, respectively. The arrows in Fig. 3(c) denote the MI transition temperatures observed in $\rho$ ($T_\rho$), which are defined as the peak position in the d[log $\rho$ / $\rho_{35°C}$] / d$T$ versus $T$. $T_\rho$ for the VO$_2$ films depends on the substrate (*i.e.*, $T_\rho$ of the VO$_2$ film on (0001) α-Al$_2$O$_3$ is 64 °C, which is close to 68 °C of VO$_2$ bulks,[1] but decreases to 59 °C on (11$\bar{2}$0) α-Al$_2$O$_3$ and 50 °C on (001) TiO$_2$). The difference in $T_\rho$ should be related to the interfacial strain effect between the films and the substrates because it has been reported that strains imposed on $VO_2$ films by substrates significantly affect $T_\rho$.[30] In general, $T_\rho$ depends on the $c$-axis length of the rutile structure in $VO_2$ because the V−V chain length along the $c$-axis plays an important role in the MI transition.[5, 31] It should be noted that the transition from an insulator to a metal in bulk $VO_2$ simultaneously induces shrinkage by 1% along the $c$-axis of rutile structure,[2,32] indicating that the lattice shrinkage would drive the electronic phase toward the metal.

To estimate the $c$-axis length, the monoclinic structure was converted into the pseudo-rutile structure. Figure 4(a) depicts the relations between $T_\rho$ and the $c$-axis lattice parameters estimated from the XRD patterns of the $VO_2$ films (Fig. 1) as well as that of bulk $VO_2$.[33] $T_\rho$ of $VO_2$ films with different crystallographic orientations monotonically decreases as the $c$-axis lattice parameters shrink in the pseudo-rutile structure compared to those of bulk $VO_2$. Because the film on (0001) α-$Al_2O_3$ should be structurally relaxed as a result of the large lattice mismatch and different crystallographic symmetry, $T_\rho$ is comparable to that of bulk $VO_2$. On the other hand, the films on (11$\bar{2}$0) α-$Al_2O_3$ and (001) $TiO_2$ are partially strained, and $T_\rho$ consequently decreases. Meanwhile, the transition temperatures observed in $S$ ($T_S$), which are determined as peak position in d$S$ / d$T$ vs. $T$ (Fig. 3(d)), also decreases as the $c$-axis length decreases (Fig. 4(b)). These results indicate that the distortion along the $c$-axis of the rutile structure is the dominant factor for reducing $T_\rho$ and $T_S$ in the present $VO_2$ films.

**IV. DISCUSSION**

To more clearly show the MI transition process, onset and offset temperatures of $T_\rho$ and $T_S$ for the transition from an insulator to a metal are denoted by open and closed triangles, respectively, in Figs. 3(c) and (d). Color bars indicate the temperature range of the MI transition width where the metallic and insulating phases should coexist. It is noteworthy that the MI transition temperature width becomes broader as $T_\rho$ and $T_S$ decrease, presumably due to the distortion of the *c*-axis length in the rutile structure of the films, which leads to an inhomogeneous domain distribution. Although the offset $T_\rho$ and offset $T_S$ are almost the same, the onset $T_\rho$ is much higher than the onset $T_S$, (*i.e.*, the transition temperature where the insulating phase begins to transform into metallic phase is higher for the $\rho$-measurement compared to that for the *S*-measurement).

One plausible reason for this difference should come from the domain distribution of the metallic and insulating phases at the nanoscale.[15] Figure 5 schematically depicts the domain configurations at each temperature (a-d) around the MI transition according to a previous report[15] and the corresponding changes in $\rho$–$T$ and $S$–$T$ as functions of temperature. As the temperature increases from (a) to (d), the majority phase changes from an insulator (a) to a metal (d), where the metallic and insulating domains coexist at (b) and (c); metallic domains first form inside the insulating phase, then some domains connect with each other and, finally, the entire region becomes metallic phase. It has been reported that $VO_2$ films show percolation conduction, depending on the domain configuration, and the $\rho$ significantly decreases when the metallic domains are connected with each other at (c),[15] *i.e.*, even though the metallic domain exists in the dominant insulating domain in (b), the contribution of the metallic domain to the conduction is extremely small due to the huge difference in resistivity between metallic and insulating phases in the order of $10^4$–$10^5$. $S$ measured in VO$_2$ film at the temperature around $T_{MI}$ should also come from the combined voltages developed across the metallic and insulating domains. Here, we roughly assume the domain distribution at (b) and (c); the metallic and insulating domains are mainly connected in series at (b), while they are linked in parallel at (c), in order to briefly discuss the effect of coexistence of metallic and insulating domains on $S$. It should be pointed out that the $S$ is complex in the multiphase materials, where the complexity originates from conductivity-weighted $S$ of the components for parallel-path arrangements and thermal conductivity for series arrangements. Thermal conductivity of VO$_2$ films increases by 60% in the metallic phase from insulating one,[34] which could be comprehended as the contribution of the increased electrical conductivity assuming that the lattice contribution of thermal conductivity remains the same across the MI transition. Although there is no such simple description for $S$ in the multiphase materials, we take no account on the difference of thermal conductivity for the discussion. The $S$ at (b) can be simply expressed by the equation of $|S| = x \cdot |S|_M + (1-x) \cdot |S|_I$ in series model, where $x$ is defined as the effective volume fraction of metallic domains against insulating one. When neglecting the contribution from the metal-insulator domain walls, the total $S$ are a sum of the contributions from the metallic and insulating domains, but in the case of $|S|$, the difference between those of the metal (23 μV K$^{-1}$) and the insulator (~200 μV K$^{-1}$) is not large, which should allow the hidden metallic phase to be detected in the majority insulating phase at (b) using $S$ measurements. Meanwhile, the $S$ at (c) can be expressed by the equation of $|S| = (x \cdot \sigma_M \cdot |S|_M + (1-x) \cdot \sigma_I \cdot |S|_I) / (x \cdot \sigma_M + (1-x) \cdot \sigma_I)$ in parallel model. Since the $S$ largely depends on the each conductivity ($\sigma$) in the parallel

circuit, the total $S$ is more sensitive to that of metallic domains when the metallic domains are connected with each other at (c). These considerations suggest that the MI transition process is much clearer when following the measurement of $S$ than that of $\rho$.

## V. CONCLUSION

In summary, we investigated the insulator-to-metal phase transition of $VO_2$ thin films with controlled lattice distortion by $S$-measurements. $VO_2$ epitaxial films with different crystallographic orientations were fabricated on (0001) α-$Al_2O_3$, ($11\bar{2}0$) α-$Al_2O_3$, and (001) $TiO_2$ substrates. MI transitions are observed both in $\rho-T$ and $S-T$, and the $T_\rho$ and $T_S$ systematically decrease, along with the broadening of the MI transition temperature width, due to the distortion of the $c$-axis length in the pseudo-rutile structure for each $VO_2$ film. Moreover, the onset $T_S$, which is where the insulating phase starts to become a metallic one, is much lower than onset $T_\rho$. This difference is attributed to the more sensitive detection of the hidden metallic domains that remain in the majority insulating phase via $S$-measurements, which cannot be detected in $\rho$-measurements. These results indicate that the $S$-measurements provide a straightforward and excellent technique to investigate the MI transition process of $VO_2$ in detail.

**ACKNOWLEDGMENTS**

H.O. was supported by Grant-in-Aid for Scientific Researches A (No. 25246023), Innovative Areas "Nano Informatics" (No. 25106007) from JSPS, and Asahi Glass Foundation. T.K. was supported by a Grant-in-Aid for Young Scientists A (No. 15H05543) from JSPS and Murata Science Foundation.

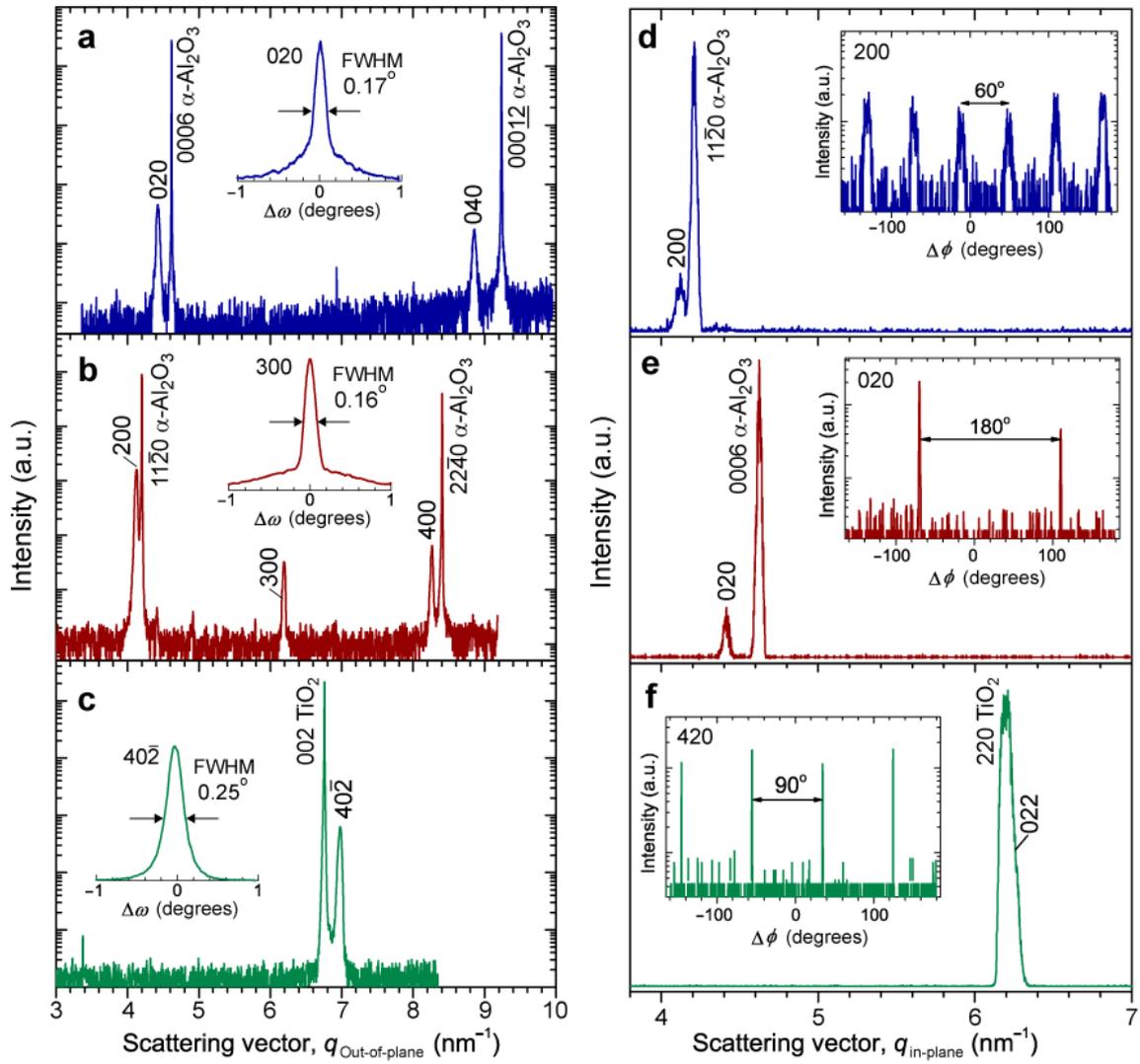

FIG. 1. (Color online) (a–c) Out-of-plane and (d–f) in-plane XRD patterns at room temperature for VO$_2$ films grown on single crystalline substrates of (0001) α-Al$_2$O$_3$ (a, d), (11$\bar{2}$0) α-Al$_2$O$_3$ (b, e), and (001) TiO$_2$ (c, f), respectively. Crystalline phases and diffraction indices are noted above the corresponding diffraction peaks. Insets of (a–c) show out-of-plane rocking curves (2θ-fixed ω-scans) and those of (d–f) show in-plane rocking curves (2θ$_\chi$-fixed φ-scans) for each diffraction peak of VO$_2$ film.

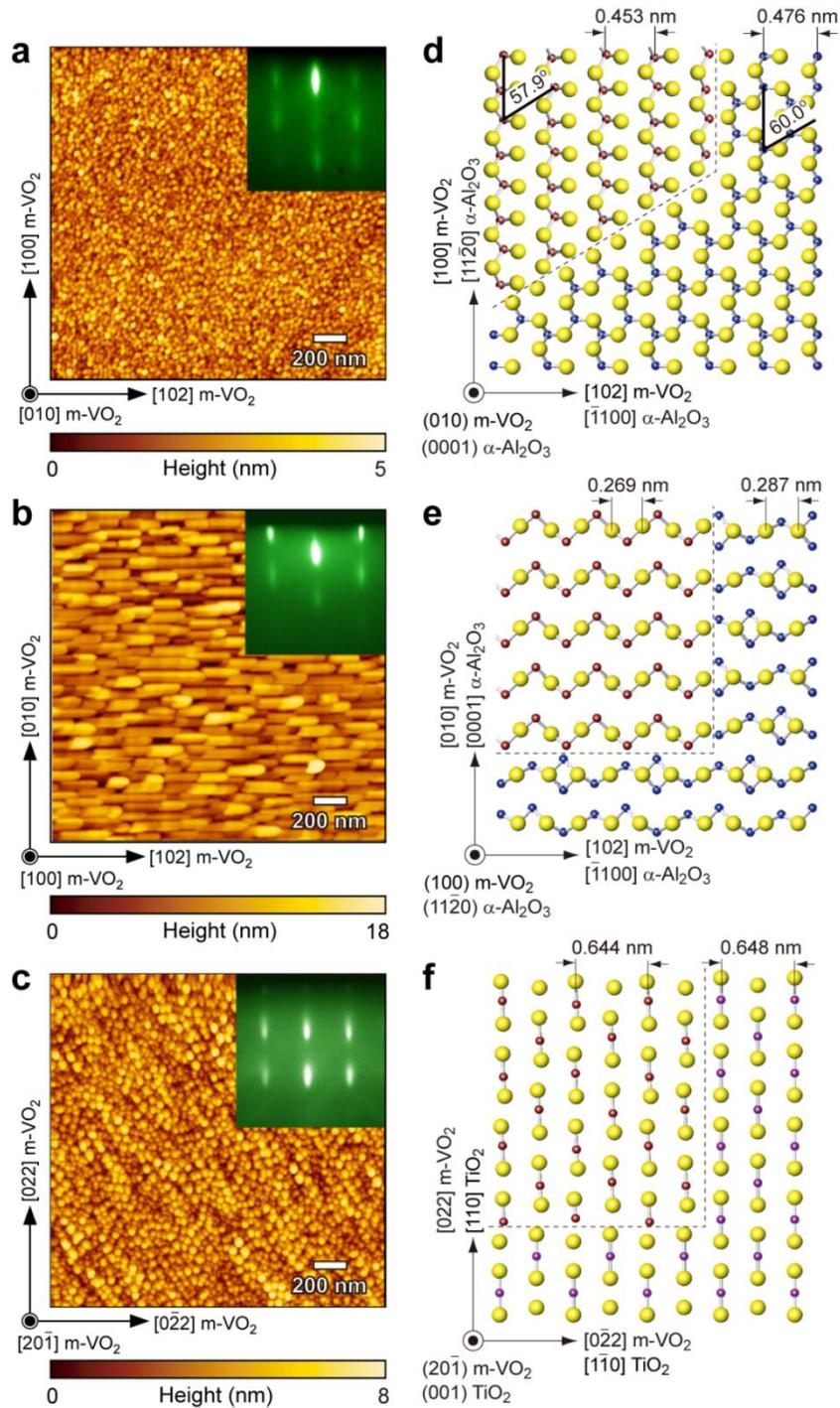

FIG. 2. (Color online) (a–c) Topographic AFM images and (d–f) schematic epitaxial relations of monoclinic (m-) $VO_2$ epitaxial films grown on substrates of (0001) α-$Al_2O_3$ (a, d), (11$\bar{2}$0) α-$Al_2O_3$ (b, e), and (001) $TiO_2$ (c, f), respectively. RHEED patterns of each $VO_2$ film are superimposed in (a–c).

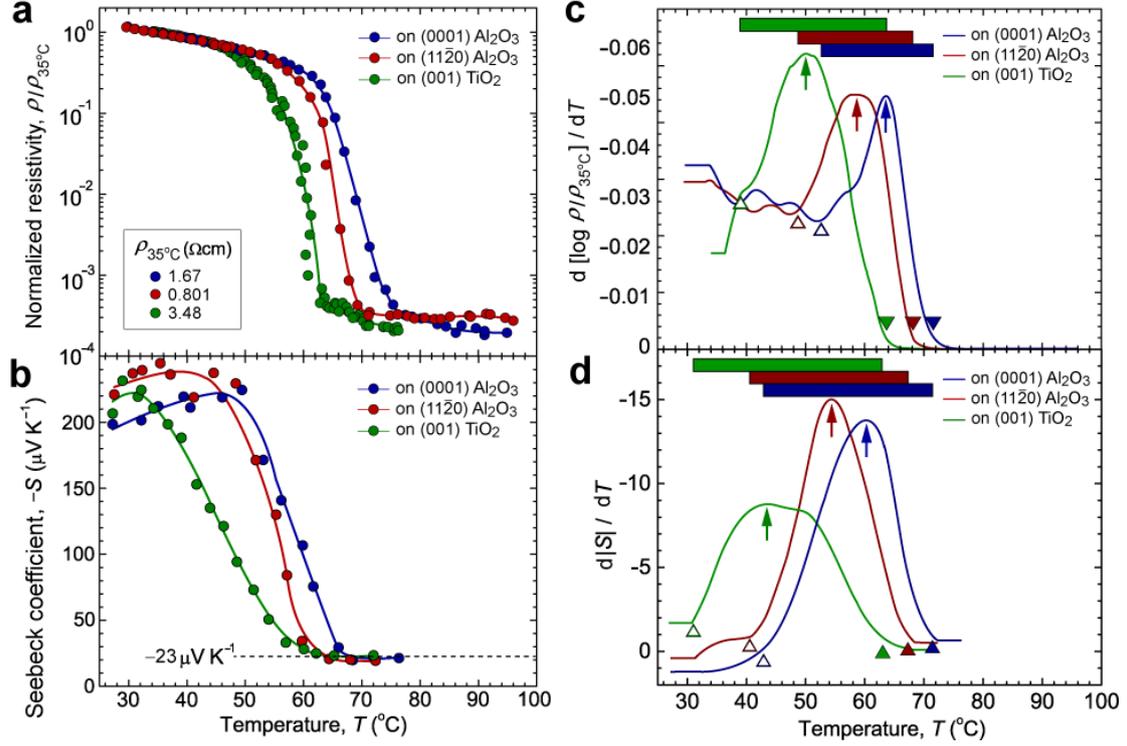

FIG. 3. (Color online) (a, b) Temperature dependences of electrical resistivity ($\rho$) and Seebeck coefficient ($S$) for VO$_2$ epitaxial films grown on (0001) α-Al$_2$O$_3$ (blue), (११$\bar{2}$0) α-Al$_2$O$_3$ (red), and (001) TiO$_2$ (green), respectively. (a) $\rho$–$T$ curves normalized by $\rho$ at 35 °C ($\rho/\rho_{35°C}$) and (b) $S$–$T$ curves measured by increasing $T$ from RT. (c, d) Temperature derivative plots of d[log $\rho/\rho_{35°C}$] / d$T$ (c) and d$S$ / d$T$ (d). The arrows indicate the positions of $T_\rho$ and $T_S$. The onset and offset temperatures of the transition from an insulator to a metal are marked by open and closed triangle symbols, respectively, and the color bars indicate the temperature range of MI transition.

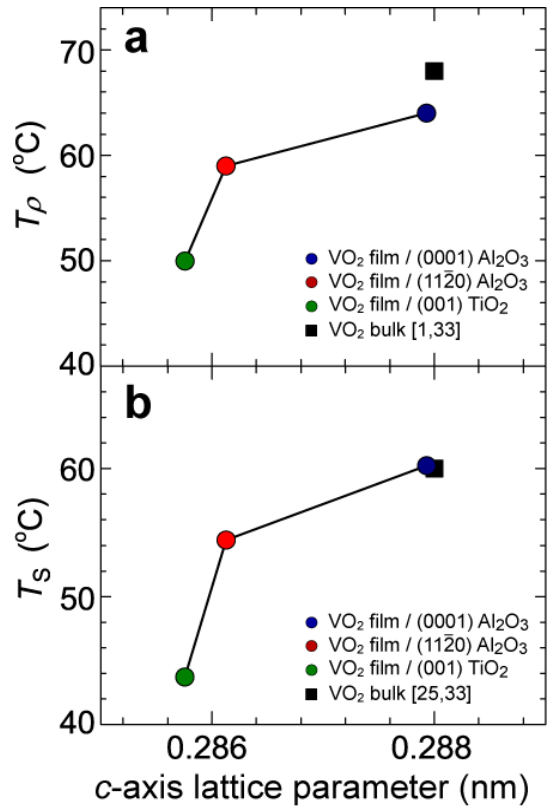

FIG. 4. (Color online) (a, b) The relations of $T_\rho$ (a) and $T_S$ (b) with estimated $c$-axis lattice parameters in the pseudo-rutile structure for $VO_2$ epitaxial films. Those of monoclinic-$VO_2$ bulk are also shown for comparison.[1, 25, 33] $T_\rho$ and $T_S$ of the $VO_2$ films decreased with shrinkage of the $c$-axis lattice parameters from those of bulk $VO_2$.

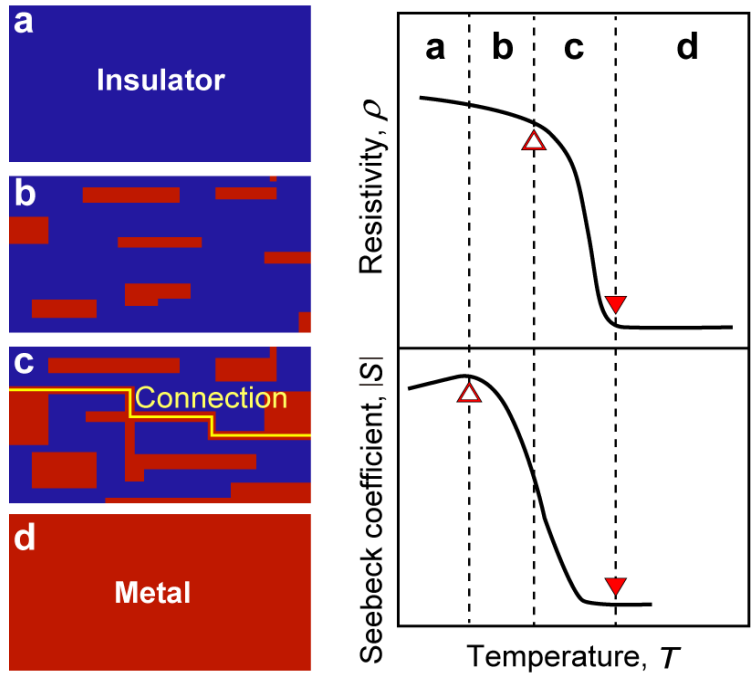

FIG. 5. (Color online) Schematic illustration of metallic and insulating domain configuration at each temperature (a–d) around MI transition,[15] and corresponding changes in $\rho-T$ and $S-T$. Majority phase changes from (a) an insulator to (d) a metal with increase of temperature, through the distribution of locally separated metallic domains (b) and connected metallic domains (c) in insulating domains. Open and closed triangle symbols indicate onset and offset temperatures of the transition from an insulator to a metal. Since $S$ can sensitively detect metallic phase remaining in majority insulating phase, which cannot be detected in resistivity, $|S|$ starts to decrease from (b).

Supplementary Material for
## "Thermopower analysis of metal-insulator transition temperature modulations in vanadium dioxide thin films with lattice distortion"

Takayoshi Katase[*], Kenji Endo, and Hiromichi Ohta[*]

*Research Institute for Electronic Science, Hokkaido University, N20W10, Sapporo 001-0020, Japan*
[*] Correspondence to: katase@es.hokudai.ac.jp, hiromichi.ohta@es.hokudai.ac.jp

**Calculation of combined Seebeck coefficients for $VO_2$ film and Cu lead**

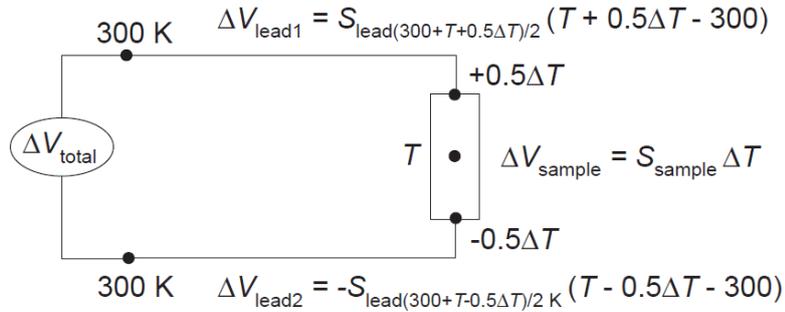

$$\Delta V_{total} = S_{lead(300+T+0.5\Delta T)/2}(T + 0.5\Delta T - 300) + S_{sample}\Delta T$$
$$- S_{lead(300+T-0.5\Delta T)/2\ K}(T - 0.5\Delta T - 300)$$

$\Delta T = 0\ K \longrightarrow \Delta V_{total} = 0\ V$

$\Delta T \fallingdotseq 0\ K \longrightarrow \Delta V_{total} \sim S_{lead(300+T)/2}\Delta T + S_{sample}\Delta T$

$S_{sample} = \Delta V_{total}/\Delta T - S_{lead(300+T)/2}$

$S_{lead(300+T)/2} \sim +1.5\ \mu V\ K^{-1} \ll S_{sample}$

$S_{sample} = \Delta V_{total}/\Delta T$